\documentstyle [12pt] {article}

\catcode`@=12
\begin{document}
\thispagestyle{empty}
\begin{flushright} UCRHEP-T275\\June 2000\
\end{flushright}
\vskip 1.0in
\begin{center}
{\Large	\bf Neutrino Masses and Leptogenesis\\}
\vskip 2.0in
{\bf Ernest Ma\\}
\vskip 0.1in
{\sl Department of Physics, University of California\\}
{\sl Riverside, California 92521, USA\\}
\end{center}
\vskip 1.2in
\begin{abstract}\
I discuss the connection between neutrino masses and leptogenesis.  I use 
three prime examples: (1) canonical seesaw, (2) triplet Higgs, and (3) 
$R$ parity violation.
\end{abstract}
\vskip 0.5in
------------------------

\noindent Talk at Third Latin American Symposium on High Energy Physics, 
Cartagena, Colombia (April 2000)

\newpage
\baselineskip 24pt

\section{Origin of Neutrino Mass}

Recent neutrino experiments \cite{1,2,3}, which are most naturally explained 
by neutrino oscillations, are strong indications that neutrinos are massive 
and mix with one another.  So where do neutrino masses come from?  In the 
minimal Standard Model, neutrinos appear only as left-handed fields in 
three electroweak doublets $(\nu_i, l_i)_L$, where $i = 1,2,3$ is the 
family index.  Charged leptons $l_i$ have right-handed components which are 
singlets, but not neutrinos.  Hence neutrinos are massless two-component 
fermions, as long as there is no physics beyond the minimal Standard Model.  
Otherwise, there may be an effective dimension-5 operator \cite{wein}
\begin{equation}
{1 \over \Lambda} (\nu_i \phi^0 - l_i \phi^+)(\nu_j \phi^0 - l_j \phi^+),
\end{equation}
where $(\phi^+, \phi^0)$ is the standard Higgs doublet and $\Lambda$ is a 
large mass, which yields a nonvanishing Majorana neutrino mass matrix as 
$\phi^0$ acquires a nonzero vacuum expectation value in the spontaneous 
breaking of the $SU(2) \times U(1)$ electroweak gauge symmetry.

All models of neutrino mass with the same low-energy particle content as 
that of the minimal Standard Model differ only in the way the above effective 
operator is realized \cite{ma}.  The most well-known such model by far is 
the canonical seesaw model \cite{seesaw}, where three right-handed neutrino 
singlets with large Majorana masses are added.  This amounts to inserting 
a heavy internal fermion line between the two singlet factors of Eq.~(1). 
The corresponding diagram can be read off to obtain the neutrino mass matrix 
as
\begin{equation}
({\cal M}_\nu)_{ij} = {f_i f_j \langle \phi^0 \rangle^2 \over m_N},
\end{equation}
where $f_i$ are the Yukawa couplings linking $\nu_i$ with the heavy singlet 
$N$ with mass $m_N$.  Obviously, we need three $N$'s to obtain three 
naturally small seesaw masses for $\nu_e$, $\nu_\mu$, and $\nu_\tau$. 
On the other hand, other mechanisms are available \cite{ma}, the simplest 
alternative being the addition of a heavy scalar triplet \cite{masa}.  
This is easily recognized if we rewrite Eq.~(1) as
\begin{equation}
{1 \over \Lambda} [ \nu_i \nu_j \phi^0 \phi^0 - (\nu_i l_j + l_i \nu_j) 
\phi^0 \phi^+ + l_i l_j \phi^+ \phi^+ ],
\end{equation}
hence an insertion of the heavy scalar triplet
\begin{equation}
\xi = (\xi^{++}, \xi^+, \xi^0)
\end{equation}
into the above yields a neutrino mass matrix given by
\begin{equation}
({\cal M}_\nu)_{ij} = {2 f_{ij} \mu \langle \phi^0 \rangle^2 \over m_\xi^2},
\end{equation}
where $f_{ij}$ are the Yukawa couplings of $\xi$ to the lepton doublets and 
$\mu$ is its coupling (with the dimension of mass) to the scalar doublets. 
Note that Eq.~(5) can also be interpreted as due to $\langle \xi^0 \rangle$, 
i.e. \cite{masa}
\begin{equation}
\langle \xi^0 \rangle = {\mu \langle \phi^0 \rangle^2 \over m_\xi^2} << m_\xi.
\end{equation}
This shows explicitly that the vacuum expectation value of a heavy scalar 
field may in fact be very small.

\section{Leptogenesis}

Given that lepton number is not conserved in models of Majorana neutrino 
masses, the attractive possibility exists that a primordial lepton 
asymmetry may be created in the early Universe, which then gets converted 
into the present observed baryon asymmetry through the $B + L$ violating, 
but $B - L$ conserving interactions of the electroweak sphalerons \cite{krs}. 
In the canonical seesaw model, this is accomplished \cite{fuya} by the decays 
of $N_i$. Being Majorana fermions, $N_{1,2}$ may decay into either $l^- 
\phi^+$ with $L = 1$ or $l^+ \phi^-$ with $L = -1$.  Consider now the 
amplitude for $N_1 \to l^+ \phi^-$.  It is the sum of 3 terms: the obvious 
tree graph, the one-loop vertex correction with
\begin{equation}
N_1 \to l^- \phi^+ \to \phi^- l^+
\end{equation}
through the exchange of $N_2$, and the one-loop self-energy correction with
\begin{equation}
N_1 \to l^- \phi^+ \to N_2 \to l^+ \phi^-.
\end{equation}
Let this amplitude be denoted by $A + iB$, where $B$ is the absorptive part, 
i.e. from putting the intermediate state $l^- \phi^+$ on the mass shell.  
Then the lepton asymmetry from the decay of $N_1$ is proportional to
\begin{equation}
|A + iB|^2 - |A^* + iB^*|^2 = 4 Im (A^*B).
\end{equation}
This means that $CP$ violation is essential and that the presence of a 
different $N$, i.e. $N_2$, in the loop is necessary for leptogenesis.

Since 1995, there has been a resurgence of activity \cite{active} in this 
topic.  Consider $N_{1,2}$ and the mass matrix linking $(\overline N_{1L}, 
\overline N_{2L}, \overline N_{1R} \overline N_{2R})$ with $(N_{1L}, N_{2L}, 
N_{1R}, N_{2R})$:
\begin{equation}
{\cal M}_N = \left[ \begin{array} {c@{\quad}c} 0 & {\cal M} \\ \tilde 
{\cal M} & 0 \end{array} \right],
\end{equation}
where
\begin{eqnarray}
{\cal M} = \left[ \begin{array} {c@{\quad}c} M_1 + H_{11} & H_{12} \\ 
H_{12} & M_2 + H_{22} \end{array} \right], ~~~ \tilde {\cal M} = \left[ 
\begin{array} {c@{\quad}c} M_1 + H_{11} & \tilde H_{12} \\ \tilde H_{12} & 
M_2 + H_{22} \end{array} \right],
\end{eqnarray}
and
\begin{eqnarray}
H_{12} &=& [ M_1 \sum_\alpha h^*_{\alpha 1} h_{\alpha 2} + M_2 
\sum_\alpha h_{\alpha 1} h^*_{\alpha 2} ] \left[ 
g_{\alpha 12}^{disp} - {i \over 32 \pi} \right], 
\\ \tilde H_{12} &=& [ M_1 \sum_\alpha h_{\alpha 1} h^*_{\alpha 2} + 
M_2 \sum_\alpha h^*_{\alpha 1} h_{\alpha 2} ] 
\left[ g_{\alpha 12}^{disp} - {i \over 32 \pi}  
\right], \\ H_{jj} &=& 2 M_j \sum_\alpha |h_{\alpha j}|^2 \left[ 
g_{\alpha jj}^{disp} - {i \over 32 \pi} \right].
\end{eqnarray}
Note that $\tilde H_{12} \neq H_{12}^*$ because of the absorptive 
contribution, hence ${\cal M}_N$ is not hermitian.  This is analogous to 
$\overline K - K$ or $\overline B - B$ mixing when the decay of the 
particles is also taken into account.  The self-energy correction to 
${\cal M}_N$ implies that mass eigenstates need not be $CP$ eigenstates, 
i.e. indirect $CP$ violation as exemplified by the $\epsilon$ parameter.  
The vertex correction to ${\cal M}_N$ implies that $CP$ may be violated in 
the decay itself, i.e. direct $CP$ violation as exemplified by the 
$\epsilon'$ parameter.  $N_1 - N_2$ oscillations may also occur.  
However, all these things happen in an expanding Universe, i.e in a dense, 
hot medium which is changing with time, hence the exact details are 
complicated and are still being worked out \cite{active}.

The primordial lepton asymmetry is generated from the decay of the lightest 
$N$, i.e. $N_1$:
\begin{equation}
\delta \simeq {G_F \over 2 \pi \sqrt 2} {1 \over (m_D^\dagger 
m_D)_{11}} \sum_{j = 2,3} Im (m_D^\dagger m_D)^2_{1j} {M_1 \over M_j},
\end{equation}
where $m_D$ is the Dirac mass matrix linking $\nu$ with $N$, and 
$M_1 << M_j$ has been assumed.  This expression \cite{factor} can then 
be used to study neutrino masses and mixing from atmospheric \cite{1} 
and solar \cite{2} neutrino oscillations and to extract information 
\cite{mix} on $N_1$.  The range $10^9 - 10^{13}$ GeV for $M_1$ is 
found to be consistent with $n_B/g_* n_\gamma \sim 10^{-10}$.  Since the 
self-energy contribution is proportional to $M_1 / (M_2-M_1)$, $\delta$ 
may be enhanced if $M_2 - M_1 << M_1$, but the limit $M_2 - M_1 \to 0$ is 
not singular \cite{sing}, as it is bounded by the decay width of $N_1$.

\section{Triplet Higgs Model}

If neutrino masses come from heavy triplet scalars \cite{masa}, then the 
mixing of $\xi_1$ and $\xi_2$ through their absorptive parts, i.e. 
self-energy contributions, leads to the physical mass eigenstates 
$\psi_1$ and $\psi_2$ which are not $CP$ eigenstates.  Their decay 
asymmetries are given by
\begin{equation}
\delta_i \simeq {Im[\mu_1 \mu_2^* \sum_{k,l} f_{1kl} f^*_{2kl}] \over 
8 \pi^2 (M_1^2 - M_2^2)} {M_i \over \Gamma_i},
\end{equation}
and for $M_{1,2} \sim 10^{13}$ GeV, realistic neutrino masses and 
leptogenesis are possible.

\section{Neutrino Masses from $R$ Parity Violation}

I now come to my third example which is the generation of neutrino masses 
through $R$ parity violation in supersymmetry \cite{hasu}.  The well-known 
superfield content of the Minimal Supersymmetric Standard Model (MSSM) is 
given by
\begin{eqnarray}
&& Q_i = (u_i, d_i)_L \sim (3,2,1/6), \\ 
&& u^c_i \sim (3^*,1,-2/3), \\ 
&& d^c_i \sim (3^*,1,1/3), \\ 
&& L_i = (\nu_i, l_i)_L \sim (1,2,-1/2), \\ 
&& l^c_i \sim (1,1,1); \\ 
&& H_1 = (\bar \phi^0_1, -\phi^-_1) \sim (1,2,-1/2), \\ 
&& H_2 = (\phi^+_2, \phi^0_2) \sim (1,2,1/2).
\end{eqnarray}
Given the above transformations under the standard $SU(3) \times SU(2) \times 
U(1)$ gauge group, the corresponding superpotential should contain in general 
all gauge-invariant bilinear and trilinear combinations of the superfields. 
However, to forbid the nonconservation of both baryon number ($B$) and lepton 
number ($L$), each particle is usually assigned a dicrete $R$ parity:
\begin{equation}
R \equiv (-1)^{3B+L+2j},
\end{equation}
which is assumed to be conserved by the allowed interactions.  Hence the 
MSSM superpotential has only the terms $H_1 H_2$, $H_1 L_i l^c_j$, 
$H_1 Q_i d^c_j$, and $H_2 Q_i u^c_j$.  Since the superfield $\nu^c_i \sim 
(1,1,0)$ is absent, $m_\nu = 0$ in the MSSM as in the minimal Standard Model. 
Neutrino oscillations \cite{1,2,3} are thus unexplained.

Phenomenologically, it makes sense to require only $B$ conservation (to make 
sure that the proton is stable), but to allow $L$ violation (hence $R$ parity 
violation) so that the additional terms $L_i H_2$, $L_i L_j l^c_k$, and 
$L_i Q_j d^c_k$ may occur.  Note that they all have $\Delta L = 1$.  Neutrino 
masses are now possible \cite{numass} with Eq.~(1) realized in two ways. 
From the bilinear terms
\begin{equation}
-\mu H_1 H_2 + \epsilon_i L_i H_2,
\end{equation}
a $7 \times 7$ neutralino-neutrino mass matrix is obtained:
\begin{equation}
\left[ \begin{array}{c@{\quad}c@{\quad}c@{\quad}c@{\quad}c} M_1 & 0 & 
-g_1 v_1 & -g_1 v_2 & -g_1 u_i \\ 0 & M_2 & g_2 v_1 & -g_2 v_2 & 
g_2 u_i \\ -g_1 v_1 & g_2 v_1 & 0 & -\mu & 0 \\ g_1 v_2 & -g_2 v_2 
& -\mu & 0 & \epsilon_i \\ -g_1 u_i & g_2 u_i & 0 & \epsilon_i & 0 
\end{array} \right],
\end{equation}
where $v_{1,2} = \langle \phi^0_{1,2} \rangle /2$ and $u_i = \langle \tilde 
\nu_i \rangle /2$, with $i = e, \mu, \tau$.  Note first the important fact 
that a nonzero $\epsilon_i$ implies a nonzero $u_i$. Note also that even 
if $u_i/\epsilon_i$ is not the same for all $i$, only one linear combination 
of the three neutrinos gets a tree-level mass \cite{chfe}.  From the 
trilinear terms, neutrino masses are also obtained, but now as one-loop 
radiative corrections.  Note that these occur as the result of supersymmetry 
breaking and are suppressed by $m_d^2$ or $m_l^2$.

\section{$L$ Violation and the Universe}

As noted earlier, the $R$ parity violating interactions have $\Delta L 
= 1$.  Furthermore, the particles involved have masses at most equal to the 
supersymmetry breaking scale, i.e. a few TeV.  This means that their 
$L$ violation together with the $B + L$ violation by sphalerons \cite{krs} 
would erase any primordial $B$ or $L$ asymmetry of the Universe \cite{erase}. 
To avoid such a possibility, one may reduce the relevant Yukawa couplings 
to less than about $10^{-7}$, but a typical minimum value of $10^{-4}$ is 
required for realistic neutrino masses.  Hence the existence of the present 
baryon asymmetry of the Universe is unexplained if neutrino masses originate 
from these $\Delta L = 1$ interactions.  This is a generic problem of all 
models of radiative neutrino masses where the $L$ violation can be traced 
to interactions occuring at energies below $10^{13}$ GeV or so.

\section{Leptogenesis from $R$ Parity Violation}

Once the notion of $R$ parity violation is introduced, there are many new 
terms to be added in the Lagrangian.  Some may be responsible for realistic 
neutrino masses and may even participate in the erasure of any primordial 
$B$ or $L$ asymmetry of the Universe, but others may be able to produce a 
lepton asymmetry on their own which then gets converted into the present 
observed baryon asymmetry of the Universe through the sphalerons.  A recent 
proposal \cite{hms} shows how this may happen in a specific model.

Consider the usual $4 \times 4$ neutralino mass matrix in the $(\tilde B, 
\tilde W_3, \tilde h_1^0, \tilde h_2^0)$ basis:
\begin{equation}
\left[ \begin{array} {c@{\quad}c@{\quad}c@{\quad}c} M_1 & 0 & -s m_3 & s m_4 
\\ 0 & M_2 & c m_3 & -c m_4 \\ -s m_3 & c m_3 & 0 & -\mu \\ s m_4 & -c m_4 & 
-\mu & 0 \end{array} \right],
\end{equation}
where $s = \sin \theta_W$, $c = \cos \theta_W$, $m_3 = M_Z \cos \beta$, 
$m_4 = M_Z \sin \beta$, and $\tan \beta = v_2/v_1$.  The above assumes 
that $\epsilon_i$ and $u_i$ are negligible in Eq.~(26), which is 
a good approximation because neutrino masses are so small.  We now choose 
the special case of
\begin{equation}
m_3, ~m_4 << M_2 < M_1 < \mu.
\end{equation}
As a result, the two higgsinos $\tilde h^0_{1,2}$ form a heavy Dirac particle 
of mass $\mu$ and the other two less heavy Majorana fermion mass eigenstates 
are
\begin{eqnarray}
\tilde B' &\simeq& \tilde B + {sc \delta r_1 \over M_1-M_2} \tilde W_3 + ..., 
\\ \tilde W'_3 &\simeq& \tilde W_3 - {sc \delta r_2 \over M_1-M_2} \tilde B 
+ ...,
\end{eqnarray}
where $\delta = M_Z^2 \sin 2 \beta / \mu$, and
\begin{equation}
r_{1,2} = {1 + M_{1,2}/\mu \sin 2 \beta \over 1 - M_{1,2}^2/\mu^2}.
\end{equation}

We now observe that whereas $\tilde B$ couples to both $\bar l_L \tilde l_L$ 
and $\bar l^c_L \tilde l^c_L$, $\tilde W_3$ couples only to $\bar l_L \tilde 
l_L$ because $l^c_L$ is trivial under $SU(2)_L$.  On the other hand, 
$R$ parity violation implies that there is $\tilde l_L - h^-$ mixing as 
well as $\tilde l^c_L - h^+$ mixing.  Therefore, both $\tilde B'$ and 
$\tilde W'_3$ decay into $l^\pm h^\mp$ and may be the seeds of a lepton 
asymmetry in such a scenario.

Let the $\tilde l_L - h^-$ mixing be very small (which is a consistent 
assumption for realistic neutrino masses from bilinear $R$ parity violation). 
Then $\tilde W'_3$ decays only through its $\tilde B$ component.  Hence the 
decay rate of the LSP (Lightest Supersymmetric Particle), i.e. $\tilde W'_3$, 
is very much suppressed, first by $\delta$ and then by the $\tilde l^c_L - 
h^+$ mixing which will be denoted by $\xi$.  Our construction is aimed at 
satisfying the out-of-equilibrium condition:
\begin{equation}
\Gamma (\tilde W'_3 \to l^\pm h^\mp) < H = 1.7 \sqrt {g_*} (T^2/M_{Pl})
\end{equation}
at the temperature $T \sim M_2$, where $H$ is the Hubble expansion rate of 
the Universe with $g_*$ the effective number of massless degrees of freedom 
and $M_{Pl}$ the Planck mass.  This implies
\begin{equation}
\left( {\xi |\delta| r_2 \over M_1-M_2} \right)^2 {1 \over M_2} < 1.9 \times 
10^{-14} {\rm GeV}^{-1},
\end{equation}
where we have used $g_* = 10^2$ and $M_{Pl} = 10^{18}$ GeV.

The lepton asymmetry generated from the decay of $\tilde W'_3$ has both 
vertex and self-energy loop contributions from the insertion of $\tilde B'$. 
However, the coupling of $\tilde B'$ to $l^\pm h^\mp$ is suppressed only by 
$\xi$ and not by $\delta$, thus a realistic asymmetry may be established if 
$\xi$ is not too small.  Let $x \equiv M_2^2/M_1^2$, then the decay 
asymmetry of $\tilde W'_3$ is given by
\begin{equation}
\epsilon = {\alpha \xi^2 \over 2 \cos^2 \theta_W} {Im \delta^2 \over 
|\delta|^2} {\sqrt x g(x) \over 1-x},
\end{equation}
where
\begin{equation}
g(x) = 1 + {2(1-x) \over x}\left[ \left( {1+x \over x} \right) \ln (1+x) - 1 
\right].
\end{equation}
The phase of $\delta$ comes from the relative phase between $M_1$ and $M_2$. 

To make sure that at $T \sim M_2$, the $L$ violating processes $l^\pm h^\mp 
\leftrightarrow l^\mp h^\pm$ through $\tilde B'$ exchange do not erase 
$\epsilon$, we require
\begin{equation}
\left( {2 e^2 \xi^2 \over \cos^2 \theta_W} \right)^2 {1 \over M_1^2} {T^3 
\over 32 \pi} {f(x) \over (1-x)^2} < H
\end{equation}
at $T \sim M_2$, where
\begin{eqnarray}
f(x) = 1 + {2(1-x) \over x^2} [(1+3x) \ln (1+x) -x(1+x)],
\end{eqnarray}
which implies
\begin{equation}
{\xi^4 \over M_2} {x f(x) \over (1-x)^2} < 2.6 \times 10^{-14} {\rm GeV}^{-1}.
\end{equation}
A sample solution is
\begin{eqnarray}
&& M_1 = 3 ~{\rm TeV}, ~{\delta \over M_1-M_2} = 8.3 \times 10^{-4}, 
\nonumber \\ && M_2 = 2 ~{\rm TeV}, ~\xi = 2 \times 10^{-3}.
\end{eqnarray}
In that case,
\begin{equation}
\epsilon = 3.6 \times 10^{-8} ~Im \delta^2/|\delta|^2,
\end{equation}
and
\begin{equation}
{n_B \over g_* n_\gamma} \sim {\epsilon \over 3 g_*} \sim 10^{-10} 
{Im \delta^2 \over |\delta|^2}.
\end{equation}
Hence realistic leptogenesis is possible if $\xi \sim 10^{-3}$ can be 
obtained.

The origin of $\tilde l^c_L - h^+$ mixing in $R$ parity violation is 
usually the term $H_1 \tilde L \tilde l^c$, which is very small because 
$\langle \tilde \nu \rangle$ has to be very small.  To obtain $\xi \sim 
10^{-3}$, we need to add the nonholomorphic \cite{nonho} term $H_2^\dagger 
H_1 \tilde l^c$ which is generally unconstrained.

\section{Conclusion}

Given a mechanism for generating small Majorana neutrino masses, it is often 
a {\it bonus} to find that leptogenesis is possible at the same time.  In 
the canonical seesaw and triplet Higgs models, the same new physics is 
responsible for both.  In $R$ parity nonconserving supersymmetry, they may 
come from different sectors of the theory.

\section*{Acknowledgments}

I thank Enrico Nardi and William Ponce and the other organizers for their 
great hospitality at Cartagena.  This work was supported in part by the 
U.~S.~Department of Energy under Grant No.~DE-FG03-94ER40837.

\bigskip

\bibliographystyle{unsrt}

\begin{thebibliography}{99}
\bibitem{factor} The numerical coefficient of $\delta$ differs from one 
another in several early references.  The one I use here was first given by 
Buchmuller and Plumacher in Ref.\cite{active}.
\end{thebibliography}

\end{document}